\documentstyle[12pt]{article}
\begin{document}
\thispagestyle{empty}
\null\vskip -1cm
\centerline{
\vbox{
\hbox{January 1996}\vskip -9pt
\hbox{hep-ph/9601326}\vskip -9pt
     }
\hfill 
\vbox{
\hbox{NHCU-HEP-96-1}
\hbox{UICHEP-TH/96-6}
     }     } \vskip 1cm

\centerline
{\large \bf 
Solutions to the $R_b$, $R_c$ and $\alpha_s$ Puzzles by Vector Fermions
} 
\vspace{1cm}
\centerline           {
Chia-Hung V. Chang$^{(1)}$ 
Darwin Chang$^{(1,2)}$,
and
Wai-Yee Keung$^{(3)}$ }
\begin{center}
\it 
$^{(1)}$Physics Department, 
National Tsing-Hua University, Hsinchu 30043, Taiwan, R.O.C.\\
\vspace{.5cm}
$^{(2)}$Institute of Physics, Academia Sinica, Taipei, R.O.C.\\
\vspace{.5cm}
$^{(3)}$Physics Department, University of Illinois at Chicago, 
IL 60607-7059, USA
\end{center}

\begin{abstract}
We propose two minimal extensions of Standard Model, both of which can
easily accommodate the recent puzzling observations about the excess in
$R_b$, the deficit in $R_c$ and the discrepancy in the low energy and
high energy determinations of $\alpha_{s}$.  Each model requires three
additional heavy vectorial fermions in order to resolve the puzzles. The
current phenomenological constraints and the new potential phenomena are
also discussed. 
\end{abstract}

\vspace{1in}
\centerline{PACS numbers: 13.38.Dg, 12.15.Ff, 12.90.+b \hfill}
\newpage

Recently it was reported\cite{lep,lep2} by LEP Collaborations that the 
measured rate of $Z \rightarrow b \bar b$  is greater than the 
prediction of Standard Model while that of $Z \rightarrow c \bar c$  
is smaller.  
This is quite significant given the impressive confirmation of the standard
model by other precision electroweak tests at the $Z^{0}$ resonance.
Given  $R_b \equiv \Gamma (Z 
\rightarrow b \bar b)/\Gamma_{had}$ and 
$R_c \equiv \Gamma (Z \rightarrow c \bar c)/\Gamma_{had}$, the discrepancies
can be summarized as
\begin{center}
\begin{math}
\begin{array} {|c|c|c|c|} \hline
{}~ & \rm Measurement & \rm SM & \rm Pull \\
\hline
R_b & 0.2219 \pm 0.0017 & 0.2156 & 3.7 \\
R_c & 0.1543 \pm 0.0074 & 0.1724 & -2.5 \\
\hline
\end{array}
\end{math}
\end{center}

\noindent 
Here SM stands for the standard-model fit with $m_t = 178$ GeV and 
$m_H = 300$ GeV, and ``Pull" is the difference between measurement and 
fit in units of the measurement error.

At the same time, the $\alpha_{s}$ problem becomes more acute with
improved precision data from $Z$ decays.
The strong coupling constant $\alpha_{s}$ extracted 
from high energy measurements at $M_Z$ seems to be larger than that from 
low energy measurements, such as deep
inelastic scattering and lattice calculations\cite{lep,hagi,shif}.
The $\alpha_{s}(M_{Z})$ calculated from the total hadronic width in Z 
decays is $0.125 \pm 0.005$\cite{olch,shif}. On the other hand, low energy
measurements all cluster around $\alpha_{s}(M_{Z}) \sim 0.11$.
It seems there is a substantial gap between the two.
Although more data in the future might eliminate these discrepancies, 
it is possible that this ``$R_{b}$--$R_{c}$'' plus $\alpha_{s}$ crisis are 
indicating the same new physics beyond the Standard Model.  

Several extensions \cite{higgs,susy,6} of the Standard Model have been 
proposed to address these puzzles.
In these models, one-loop corrections to the $Z b \bar b$ vertex 
from the non-standard sector will enhance the $b$ quark partial width. 
With the hadronic total width also enhanced, the QCD 
corrections needed to fit the observed total width is reduced.
Thus the observed data point to a smaller $\alpha_{s}$ than that in 
the Standard Model, as favored by low energy measurements. 
However, these attempts all fail to account for the large $R_{c}$ deficit.
In addition, the first two scenarios might be in 
potential conflict with top quark decay\cite{7}.

More recently, two papers\cite{ma,wshou} pointed a new direction in the 
extensions of Standard Model which may provide a simple solution to the 
above discrepancies.  Both papers suggested to 
resolve the discrepancies by introducing new vectorial fermions that mixes 
with $b$ and/or $c$ quarks.  The mixing will reduce or enhance the 
couplings of 
the quarks to $Z$ boson depending on the gauge quantum numbers of the new 
fermions. We shall call this class of solutions `vectorial fermionic solutions'
to the puzzles. In Ref.\cite{wshou}, only a vectorial pair of singlet is 
introduced to reduce the partial width of $c\bar{c}$. This could solve the 
$R_c$ puzzle while leaving the $R_b$ puzzle only slightly ameliorated. 
On the other hand, Ref.\cite{ma}, a vectorial pair of singlet plus 
a vectorial pair of triplet are added to resolve both puzzles at the 
same time at tree level. As a price of solving both problems, 
Ma's model also reduces the prediction for the total hadronic width
$\Gamma_{had}$ and thus renders a surplus in the observed leptonic
branching ratio $R_{l} \equiv \Gamma_{had}/\Gamma_{l}$, which can not be
accommodated by assuming a smaller $\alpha_{s}$. 
In this paper we propose and analyze two minimal extensions of Standard 
Model 
which are nevertheless sufficient to resolve the $R_{b}$ and $R_{c}$ 
puzzles and simultaneously lower the value of $\alpha_{s}$ extracted from 
$Z$ decay.  In the first minimal extension, only a  vectorial triplet of 
fermions are needed while; while in the  second one, one 
needs a vectorial singlet plus a vectorial doublet of fermions.  The 
first model involves less parameter and therefore has greater predicting 
power.

We shall start by analyzing the fermion mixing in the general context and 
then demonstrate that our resulting models are indeed the simplest ones of 
the class.

In general, the coupling of $Z$ meson with fermions can be written as 
\begin{equation} 
\frac{g}{\cos \theta_W}  Z^{\mu} 
(\, g^{f}_L \bar{f}_L\gamma_{\mu}f_L + g^{f}_R 
\bar{f}_R\gamma_{\mu}f_R \, )
\end{equation}
where 
\begin{equation}
g^{f}_{L,R} =  T_{fL,R}^{3} - Q_{f} \sin^{2} \theta_W 
\;.
\end{equation}
The coupling only depends on 
the weak isospin $T^{3}$ and electric charge $Q$ of the fermion. 
Thus mixing with heavy fermions of different weak isospin $T^{3}$ could 
change the coupling of quarks with $Z$ meson and the $Z$ decay partial 
width. Take the partial width into $b\bar{b}$ as an example.
Assume that there is a heavy fermion $x$ of charge $-{1\over3}$ and it
mixes with quark $b$, as well as $d$ and $s$. 
Denote the mixing matrix among left-handed (right-handed) particles as 
$U_{L}$ ($U_{R}$), which transforms mass eigenstates into gauge 
eigenstates. 
We shall specify the weak gauge eigenstates by fields with primes,
while those without primes are the mass eigenstates.
\begin{equation}
\left(  \begin{array}{c}
        d' \\
        s' \\
        b' \\
        x' 
        \end{array}  \right)_{L,R}
 =  \left( \begin{array}{cccc} 
      U_{dd} & U_{ds} & U_{db} & U_{dx} \\
      U_{sd} & U_{ss} & U_{sb} & U_{sx} \\
      U_{bd} & U_{bs} & U_{bb} & U_{bx} \\
      U_{xd} & U_{xs} & U_{xb} & U_{xx} 
       \end{array} \right)_{L,R}
       \left(  \begin{array}{c}
        d \\
        s \\
        b \\
        x
        \end{array}  \right)_{L,R}
\;.
\end{equation}
The coupling between mass eigenstate $b_{L}$ and $Z^{0}$ would become
\begin{equation}
g_{L}^{b} = [\, T^{3}_{dL}  |U_{db}|_{L}^{2}+ T^{3}_{sL} |U_{sb}|_{L}^{2} + 
T^{3}_{bL}  |U_{bb}|_{L}^{2} + T^{3}_{xL} |U_{xb}|_{L}^{2}
- Q \sin^{2} \theta_W  ]
\;,
\end{equation}
while $g_{R}^{b}$ equals a similar expression with the subscript $L$
replaced by $R$. Because the mixing matrix $U$ is unitary and quark
$d,b,s$ share the same weak isospin $T^{3}$, $g^{b}$ can be written as 
\begin{equation}
g^{b}_{L,R} = T^{3}_{bL,R} + (T^{3}_{xL,R} - T^{3}_{bL,R}) 
|U_{xb}|_{L,R}^{2} - Q \sin^{2} \theta_W  
\;.
\end{equation}
The $Z$ partial decay width into $b\bar{b}$ is proportional to 
$|g^{b}_{L}|^{2}+|g^{b}_{R}|^{2}$.
\begin{equation}
 [\, T^{3}_{bL} + (T^{3}_{xL} - T^{3}_{bL}) |U_{xb}|_{L}^{2} - 
Q \sin^{2} \theta_W \, ]^{2} +
 [\, T^{3}_{bR} + (T^{3}_{xR} - T^{3}_{bR}) |U_{xb}|_{R}^{2} - 
Q \sin^{2} \theta_W \, ]^{2}
\;.
\end{equation}
It is different from that in Standard Model.
Whether the new fermion will enhance or reduce the partial width  
depends on its weak isospin $T^{3}$.

Now it is easy to see that we can reduce $\Gamma_{c\bar{c}}$ by adding 
a left-handed singlet of $T^{3}=0$ that mixes with 
$c_{L}$\cite{ma,wshou}.
To increase $\Gamma_{b\bar{b}}$, a $T^{3} = -1$ left-handed
fermion can be introduced to enhance $|g_{L}^{b}|^{2}$.
A less obvious way is to mix $b_{R}$, which is of $T^{3}=0$, with a 
heavy right handed doublet of $T^{3}={1\over2}$. 

Next we shall show two minimal extensions of Standard Model in which
vectorial fermions with the above properties are introduced to resolve
both the $R_{b}$ and $R_{c}$ puzzles simultaneously. These are the
simplest models to accomplish that, with the smallest number of new
particles, three species of vectorial fermions in both cases. We
consider only adding vectorial fermions since anomalies are cancelled
automatically and these fermions could be heavy naturally. 

In the first model, only one vectorial triplet is needed.
The $T^{3}=0$ component will reduce 
$\Gamma_{c\bar{c}}$ and the $T^{3}=-1$ component will enhance 
$\Gamma_{b\bar{b}}$.
The triplet $Y$ can be written as 
\[
Y_{L,R} = \left( \begin{array}{c}
                 y_{1}^{5/3}   \\
                 y_{2}^{2/3}   \\
                 y_{3}^{-1/3}  \end{array} \right)_{L,R}
\;,
\]
with a gauge invariant mass term 
$M_{Y} \bar{Y}_{L} Y_{R}$.  
The mixing is induced by Yukawa couplings between the triplet $Y_{R}$ and
left-handed quark doublets.
\begin{eqnarray}
& & \xi_{3}[\, \bar{y}_{1R} t'_{L} \phi^{+} + \bar{y}_{2R} ( t'_{L} 
\phi^{0} + b'_{L} \phi^{+} ) / \sqrt{2} + \bar{y}_{3R} b'_{L} \phi^{0}\,] 
\;, \nonumber \\ 
& & \xi_{2} [\, \bar{y}_{1R} c'_{L} \phi^{+} + \bar{y}_{2R} ( c'_{L}
\phi^{0} + s'_{L} \phi^{+} ) / \sqrt{2} + \bar{y}_{3R} s'_{L} \phi^{0}\,] 
\;,\nonumber
\\ & & \xi_{1} [\, \bar{y}_{1R} u'_{L} \phi^{+} + \bar{y}_{2R} ( 
u'_{L}
\phi^{0} + d'_{L} \phi^{+} ) / \sqrt{2} + \bar{y}_{3R} d'_{L} \phi^{0}\,]
\;.
\label{yukawa} 
\end{eqnarray}
In addition we have the ordinary Yukawa couplings in the Standard Model.

$y_{3}$ mixes with the down quarks $d,s,b$.
We'll use the biunitary transformation to diagonalize 
the $3 \times 3$ mass matrix between $d,s,b$.
The mass matrix between $D'_{L} \equiv (d'_{L},s'_{L},b'_{L},y'_{3L})$ and 
$D'_{R} \equiv (d'_{R},s'_{R},b'_{R},y'_{3R})$ then become
\begin{equation}
(\bar{d}'_{L},\bar{s}'_{L},\bar{b}'_{L},\bar{y}'_{3L})
\left( \begin{array}{cccc} 
      m_{d} & 0 & 0 & \xi_{1} v \\
      0 & m_{s} & 0 & \xi_{2} v \\
      0 & 0 & m_{b} & \xi_{3} v \\
      0 & 0 & 0 & M_{Y} 
       \end{array} \right)
\left(  \begin{array}{c}
        d'_{R} \\
        s'_{R} \\
        b'_{R} \\
        y'_{3R} 
        \end{array}  \right)  \equiv \bar{D}'_{L} M_{d} D'_{R}
\;.
\end{equation}
$M_{d}$ can be written as 
\begin{equation}
M_{d} = \left( \begin{array}{cc}
               \tilde{M}_{d} & J \\
               0 & M_{Y} \end{array} \right)
\;,
\end{equation}
with $\tilde{M}_{d}$ a $3 \times 3$ matrix, which is diagonal here and 
$J$ is a $3 \times 1$ column.
It is {\em natural} to assume that the gauge invariant $M_{Y}$ is much 
larger than 
all the other elements of the matrix. The diagonalization then takes a 
simple form.
The mixing matrix $U_{L}$ ($U_{R}$) is the matrix that diagonalize 
$M_{d} M_{d}^{\dagger}$ ($M^{\dagger}_{d} M_{d}$).
$M^{\dagger}_{d} M_{d}$ has only one large element at the lower right
corner with all the other element suppressed by $(m/M_{Y})^{2}$.
Thus the mixing of $y_{3R}$ with $d_{R}, s_{R}, b_{R}$, $U_{yi}^{R} 
(i=d,s,b)$, is also suppressed by $(\xi v/M_{Y})^{2}$ and negligible.
This is a result of the fact that we cannot construct mixing Yukawa couplings 
between the triplet $y_{3L}$, the singlet $q_{R}$ and the doublet Higgs 
boson. The mixing between $y_{3L}$ and $b_{L},s_{L},d_{L}$ is more important.
Write $U_{L}$ as 
\begin{equation}
U_{L} = \left( \begin{array}{cc}
               K & R \\
               S & T \end{array} \right)
\;.
\end{equation}
$K,R,S$ are respectively a $3 \times 3$ matrix, an $1 \times 3$ column
and a $3 \times 1$ row and $T$ a number.
The various elements can be solved in the large $M_{Y}$ 
approximation\cite{lavoura}.
In this approximation, $T$ is equal to one.
$K$ equals the unitary matrix that diagonalizes 
$\tilde{M}_{d}\tilde{M}^{\dagger}_{d}$, which is just unity matrix in 
this case.
The column $R$ and row $S$ can also be calculated
\begin{equation}
R = \frac{1}{M_{Y}} J, \hspace{2em} S = -\frac{1}{M_{Y}} J^{\dagger} K
= -\frac{1}{M_{Y}} J^{\dagger}
\;.
\end{equation}
The mixing of $d_{iL}$ with $y_{3L}$, ie.\ $R_{i},S_{i}$ is 
approximately $\xi_{i} v / M_{Y}$.
As a result, $\Gamma_{b\bar{b}}$ is proportional to
\begin{equation}
 \left( - \frac{1}{2} - \frac{1}{2} |S_{3}|^{2} + 
   \frac{1}{3} \sin^{2} \theta_W \right)^{2} 
+ \left( \frac{1}{3} \sin^{2} \theta_W\right)^{2} 
\;,
\end{equation}
with
\begin{equation}
S_{3} = -\frac{\xi_{3}v}{M_{Y}}
\;.
\end{equation}
To fit the observed $R_{b}$, we need
\begin{equation}
|S_{3}|^{2} = \left(\frac{\xi_{3}v}{M_{Y}}\right)^{2} = 0.0127 \pm 0.0034
\;.
\end{equation}

The charge ${2\over3}$ quarks will mix with $y_{2}$ with the mass matrix 
\begin{equation}
M_{u} = \left( \begin{array}{cc}
               \tilde{M}_{u} & J \\
               0 & M_{Y} \end{array} \right)
\;.
\end{equation}
Note that $J$ is identical to the same column in $M_{d}$. 
However $\tilde{M}_{u}$ is no longer diagonal in this basis. 
$\tilde{M}_{u}\tilde{M}^{\dagger}_{u}$ is diagonalized by the KM 
matrix $V_{KM}$. 
Denote the mixing matrix that diagonalizes 
$M_{u}M^{\dagger}_{u}$ as $U'_{L}$
\begin{equation}
U_{L}' = \left( \begin{array}{cc}
               K' & R' \\
               S' & M_{Y} \end{array} \right)
\;.
\end{equation}
In the heavy $M_{Y}$ approximation,
\begin{equation}
K' \approx V_{KM}^{\dagger}
\;,
\end{equation}
and 
\begin{equation}
S' = -\frac{1}{M_{Y}} J^{\dagger} V_{KM}^{\dagger}
\end{equation}
The partial width $\Gamma_{c\bar c}$ is proportional to
\begin{equation}
 \left(  \frac{1}{2} - \frac{1}{2} |S'_{2}|^{2} 
       - \frac{2}{3} \sin^{2} \theta_W
 \right)^{2} 
 + \left(-\frac{2}{3}\sin^{2} \theta\right)^{2} 
\;,
\end{equation}
It is smaller than the corresponding value when there is no mixing.
To fit the data, we need
\begin{equation}
|S'_{2}|^{2} = 0.045 \pm 0.019
\;.
\label{e:s2}
\end{equation}
Note that $S'$ is related to $S$ through the KM matrix $V_{KM}$.
There is no separate parameter for the charge ${2\over3}$ quarks.
The mixing is totally fixed by three parameters $\xi_{1,2,3}$ and 
the heavy fermion mass $M_{Y}$.

In the second model, we shall introduce a vectorial doublet and a vectorial 
singlet. 
The singlet, call it $x$, has charge ${2\over3}$ and will reduce 
$R_{c}$ just like $y_{2}$ in the first model.
Choose $x_{R}$ to be the only right-handed fermion with the gauge invariant 
mass term $M_{x} \bar{x}_{L} x_{R}$ with $x_{L}$. 
The mass mixing is induced by Yukawa couplings
$\xi_{i} \bar{q}_{Li} \tilde{\phi} x_{R}$ for $i=1,2,3$.
The analysis is the same as $y_{2}$ in the first model.

The doublet will have the weak hypercharge $-{5\over3}$,
\[
\Psi_{L,R} = \left( \begin{array}{c}
                 \Psi_{1}^{-1/3}   \\
                 \Psi_{2}^{-4/3}
                  \end{array} \right)_{L,R}
\;,
\]
with a gauge invariant mass term $M_{\Psi} \bar{\Psi}_{L} \Psi_{R}$.
The Yukawa coupling between $\Psi$ and ordinary quarks are
\begin{equation}
\xi'_{3} \bar{\Psi}_{L} \tilde{\phi} \, b_{R}
+ \xi'_{2} \bar{\Psi}_{L} \tilde{\phi} \, s_{R}
+ \xi'_{1} \bar{\Psi}_{L} \tilde{\phi} \, d_{R}
\;.
\end{equation}
The coefficients $\xi'$'s need not be the same as the $\xi$'s for
the singlet $x$. 
Therefore more parameters are involved in the second model.
The mass matrix is
\begin{equation}
(\bar{d}'_{L},\bar{s}'_{L},\bar{b}'_{L},\bar{\Psi}'_{1L})
\left( \begin{array}{cccc} 
      m_{d} & 0 & 0 & 0 \\
      0 & m_{s} & 0 & 0 \\
      0 & 0 & m_{b} & 0 \\
      \xi_{1} v & \xi_{2} v & \xi_{3} v & M_{\Psi} 
       \end{array} \right)
\left(  \begin{array}{c}
        d'_{R} \\
        s'_{R} \\
        b'_{R} \\
        \Psi'_{1R} 
        \end{array}  \right)  \equiv \bar{D}'_{L} M_{d} D'_{R}
\;.
\end{equation}
Contrary to the previous case, the mixing between $\Psi_{L}$ with 
$d_{L},b_{L},s_{L}$ is suppressed by $(m/M_{\Psi})^{2}$.
The reason is that we cannot construct mixing Yukawa couplings among
the three doublets $\Psi'_{1R}$, $q'_{L}$ and Higgs boson.
Now the mixing of $\Psi_{R}$ with $d_{R},b_{R},s_{R}$ is of the order
$\xi v/M_{\Psi}$.
$\Gamma_{b\bar{b}}$ is proportional to
\begin{equation}
 \left(-\frac{1}{2} +   \frac{1}{3}  \sin^{2} \theta_W \right)^{2} + 
 \left[\frac{1}{2} \left( \frac{\xi_{3}v}{M_{\Psi}} \right)^{2} 
      + \frac{1}{3} \sin^{2} \theta_W\right]^{2} 
\;.
\end{equation}
To fit the data, we need
\begin{equation}
  \left( \frac{\xi_{3}v}{M_{\Psi}}\right)^{2} = 0.059 \pm 0.016
\;.
\end{equation}

Generally, by assuming no non-standard Higgs boson in the theory,
if the vectorial fermions are triplets or singlets, the 
effects on $g_L$ will dominate because in such case $U_R$ is much smaller 
than $U_L$.  The singlet with the $b$-quark charge will only reduce $g^b_L$ 
which is in the wrong direction, while the singlet with the $c$-quark charge 
will reduce $g^c_L$ as data require.  In the case of a triplet, it 
can either increase or reduce both $g^b_L$ or $g^c_L$ depending on the 
hypercharge.  

On the other hand, if the new vectorial fermions are doublet the effects
on $g_R$ will dominate while those on $g_L$ are largely unchanged.  To
increase $g^b_R$ we need the new vectorial ``down-type'' quark to have
$T^3=+{1\over2}$.  To reduce $g^c_R$ we also need the new vectorial 
``up-type'' quark to have $T^3=+{1\over2}$. 

From these arguments, it is straightforward to show that the two models
we have are the ones with a minimal number (three) of new vectorial
fermions.  If one allows four new vectorial fermions, there are also two
interesting models that can be considered.  One of them is adopted by Ma
in Ref.\cite{ma} and the other model uses two vectorial doublets: one
doublet  with  $Y=-{5\over3}$ to increase $g^b_L$ and another doublet
with $Y={1\over3}$ to reduce $g^c_L$. We shall not discuss these
non-minimal models in details.

In these vectorial fermion models, tree-level
flavor-changing-neutral currents (FCNC) will in general 
arise since quarks mix with fermions of different weak 
isospin.
Next we shall analyze the FCNC constraints, especially from the kaon decays.

Because of GIM mechanism, there will be no FCNC if the heavy vectorial fermions
have the same weak isospin $T_{3}$ as the quarks they mix with.
In the first model, the only component in the neutral current that will 
generate tree level FCNC in the Kaon decay is
$- \frac{1}{2} \bar{y}'_{3L} \gamma^{\mu} y'_{3L}$.
It will give rise to a FCNC vertex involving mass eigenstates $d$ and $s$.
\begin{equation}
   - \frac{1}{2} (U_L)^{\ast}_{41} (U_L)_{42}  \bar{d}_{L} 
     \gamma^{\mu} s_{L}
\;,
\end{equation}
with
\begin{equation}
(U_L)_{41} = S_{1}  \sim  \frac{\xi_{1} v}{M_{Y}} \;, \hspace{2em}
(U_L)_{42} = S_{2}  \sim  \frac{\xi_{2} v}{M_{Y}} \sim S'_2 \;.
\end{equation}
Here $\xi_{2} v /M$ is fixed by $R_{c}$ to be about $0.2$ from
Eq.(\ref{e:s2}). Thus the coefficient of the FCNC vertex 
\begin{equation}
-{g\over2\cos\theta_{W}} \bar{d}_{L} \gamma^{\mu} s_{L} 
\end{equation}
is of the order
$ 0.2 \times \xi_{1}v/M_{Y}$.
The kaon decay  
$K_{L} \rightarrow \mu^{+} \mu^{-}$ restricts this coefficient 
to be $<3.1 \times 10^{-5}$\cite{lavoura2}.
Take $M_{Y}\sim 200{\rm GeV}$ as an illustration.
The bound for $\xi_{1}v$ is $\xi_{1}v < 32 {\rm MeV}$.
Given the $d$ quark mass of about 10 MeV, the constraint is still
quite natural. If the vectorial fermion is heavier, the constrain will be 
even looser. 
In the second model, there are more parameters involved.
The Kaon FCNC constraints will now impose a limit on $\xi_{1} \xi_{2}$.
But now $\xi_{2}$ is no longer fixed by $R_{c}$ fitting, which is related to 
$\xi'_{2}$ instead.

In our models, the strong coupling constant extracted from $R_{l}$ is
different from Standard Model. While $\Gamma_{b\bar{b}}$ is enhanced and
$\Gamma_{c\bar{c}}$ reduced as the experiment indicates,
$\Gamma_{u\bar{u}}$, $\Gamma_{d\bar{d}}$ and $\Gamma_{s\bar{s}}$ change
as well. In the first model, the changes in $\Gamma_{u\bar{u}}$ and
$\Gamma_{d\bar{d}}$ are small. Their mixing with $Y$ is determined
mainly by $\xi_{1}$, which is constrained by Kaon FCNC limit. However
the enhancement in $\Gamma_{s\bar{s}}$ is quite sizable. It is dictated
by the necessary change in $\Gamma_{c\bar{c}}$ to fit $R_{c}$. Overall,
$\Gamma_{had}$ without QCD corrections is enhanced. Thus the
$\alpha_{s}(M_{Z})$ extracted from $R_{l}$ using our first model is
smaller than using Standard Model since a smaller $\alpha_{s}$ gives
smaller QCD enhancement corrections\cite{soper}. Note that $R_{b}$ and
$R_{c}$ are insensitive to $\alpha_{s}$. In our model, the
$\alpha_{s}(M_{Z})$ calculated from $R_{l}$ is $0.098 \pm 0.007 \pm
0.005$, with the first error coming from the uncertainty in the mixing.
It is now consistent with the low energy measurements. In the second
model, $\Gamma_{s}$ is enhanced just like in the first one. Thus the
$\alpha_{s}$ extracted from $R_{l}$ will be smaller than $0.125$. 

In contrast, Ma's model omits the mixing between the heavy fermion and the
$s$ quark. Thus $\Gamma_{had}$ is reduced since the absolute deviation
of $R_{c}$ is larger than that of $R_{b}$. Extracted from a smaller
prediction for $R_{l}$, the strong coupling constant becomes
0.18\cite{bamert}, even higher than the original high value of $0.125$.
This led Ma in his paper\cite{ma} to assign the heavy fermion a
relatively small mass $M_{x}<72$GeV so as to open a new channel for the 
$Z$ boson decaying into this heavy fermion. In our models, the
$\alpha_{s}$ puzzle is resolved because of the enhancement in $R_{s}$.
Experimentally it may be a challenge to measure this $R_s$ effect. If this
could be done, it will be the most direct test of our model. 

The forward and backward asymmetries $A_{FB}^{b}$ and $A_{FB}^{c}$ will
also be affected by the mixing. The prediction in the first model agrees
well with the experimental measurements, as shown by Ma\cite{ma}. In the
second model, the asymmetry $A_{FB}^{b}$ is different from the first
one. It is equal to 0.0980, which agrees with the observed value $0.0997
\pm 0.0031$. 

One may wonder how these new vectorial fermions can be accommodated in a 
Grand Unified Theory.  The vectorial triplet $Y$ can be found is 
$(3,1,15)$ of $SU(2)_L \times SU(2)_R \times SU(4)$ which in turns can be 
found in $210$ multiplet of $SO(10)$.  The vectorial doublet $\Psi$ is a 
bit harder to accommodate.  It can be found in $(2,1,20) + 
(2,1,\overline{20})$
$SU(2)_L \times SU(2)_R \times SU(4)$ which in turns can be found in $144 
+ \overline{144}$ of $SO(10)$ or $650$ of $E_6$.

We would like to thank Bing-Ling Young, C.N. Leung and V. Barger for
useful discussions. This work is supported in part by the National Science
Council of Republic of China under grants NSC84-2112-M007-042,
NSC84-2112-M007-016 (for D.C.) and in part by U.S. Department of Energy
(for W.-Y. K.).

\end{document}